\documentclass[iop,apj]{emulateapj}
\usepackage{natbib}
\usepackage{amsfonts}
\usepackage{amssymb}

\shorttitle{Large-angle correlation and odd-parity preference in CMB}
\shortauthors{Jaiseung Kim and et al.}

\begin{document}
\title{Lack of angular correlation and odd-parity preference in CMB data}
\author{Jaiseung Kim and Pavel Naselsky}
\affil{Niels Bohr Institute \& Discovery Center, Blegdamsvej 17, DK-2100 Copenhagen, Denmark}
\email{jkim@nbi.dk}
%\submitted{Submitted to the Astrophysical Journal} 

\begin{abstract}
We have investigated the angular correlation in the recent CMB data.
In addition to the known large-angle correlation anomaly, we find the lack of correlation at small angles with high statistical significance.
We have investigated various non-cosmological contamination and additionally WMAP team's simulated data. However, we have not found a definite cause.
In the angular power spectrum of WMAP data, there exist anomalous odd-parity preference at low multipoles.
Noting the equivalence between the power spectrum and the correlation, we have investigated the association between the lack of large-angle correlation and the odd-parity preference. From our investigation, we find that the odd-parity preference at low multipoles is, in fact, a phenomenological origin of the lack of large-angle correlation. 
Futher investigation is required to find out whether the origin of the anomaly is cosmological or due to unaccounted systematics. 
The data from Planck surveyor, which has systematics distinct from the WMAP, will greatly help us to resolve its origin.
\end{abstract}


\section{Introduction}
Over the past years, there have been great successes in measurement of CMB anisotropy by ground and satellite observations \citep{WMAP7:basic_result,ACBAR2008,QUaD2,Planck_mission}.
Since the release of the WMAP data \citep{WMAP3:temperature,WMAP5:basic_result,WMAP7:basic_result}, there have been reports on various anomalies \citep{cold_spot1,cold_spot2,cold_spot_wmap3,cold_spot_origin,Tegmark:Alignment,Multipole_Vector1,Multipole_Vector2,Multipole_Vector3,Axis_Evil,Axis_Evil2,Axis_Evil3,Universe_odd,Park_Genus,Chiang_NG,correlation_Copi1,correlation_Copi2,Hemispherical_asymmetry,power_asymmetry_subdegree,Power_Asymmetry5,odd,odd_origin,lowl_anomalies,odd_bolpol,WMAP7:anomaly,odd_tension}. 
In particular, there are reports on the lack of angular correlation at large angles, which are observed in COBE-DMR data and subsequently in WMAP data \citep{correlation_COBE,WMAP1:Cosmology,correlation_Copi1,correlation_Copi2,lowl_anomalies,lowl_bias}. 
In order to figure out the cause of the anomaly, we have investigated non-cosmological contamination and additionally WMAP team's simulated data.
However, we have not found a definite cause, which makes us believe the anomaly is produced by unknown systematics or may be, indeed, cosmological.

In the angular power spectrum of WMAP data, there exist anomalous odd-parity preference at low multipoles \citep{Universe_odd,odd,odd_origin,odd_bolpol,odd_tension}.
Noting the equivalence between power spectrum and correlation, we have investigated the association between the odd-parity preference and the lack of large-angle correlation.
From our investigation, we find that the odd-parity preference at low multipoles is, in fact, a phenomenological origin of the lack of the large-angle correlation.
Even though it still leaves the fundamental question on its origin unanswered, the association between seemingly distinct anomalies will help the investigation on the underlying origin, whether cosmological or unaccounted systematics.

The outline of this paper is as follows.
In Section \ref{CMB}, we briefly discuss the statistical properties of CMB anisotropy.
In Section \ref{wmap}, we investigate the angular correlation anomalies of WMAP data, and show the lack of correlation at small angles in addition to that at large angles.
In Section \ref{systematics}, we investigate non-cosmological contamination and WMAP team's simulated data.
In Section \ref{odd}, we show the odd-parity preference at low multipoles is a phenomenological origin of the lack of the large-angle correlation.
In Section \ref{discussion}, we summarize our investigation. 

\section{Angular correlation of CMB anisotropy}
\label{CMB}
CMB anisotropy over a whole-sky is conveniently decomposed in terms of spherical harmonics: 
\begin{eqnarray}
T(\hat{\mathbf n})&=&\sum_{lm} a_{lm}\,Y_{lm}(\hat{\mathbf n}),\label{T_expansion}
\end{eqnarray}
where $a_{lm}$ and $Y_{lm}(\hat {\mathbf k})$ are a decomposition coefficient and a spherical harmonic function. 
In most of inflationary models, decomposition coefficients of CMB anisotropy follow the Gaussian distribution of the following statistical properties:
\begin{eqnarray}
\langle a_{lm}a^*_{l'm'}\rangle = \delta_{ll'} \delta_{mm'}\,C_{l} \label{C_lm}
\end{eqnarray}
where $\langle \ldots \rangle$ denotes the average over an ensemble of universes, and
$C_l$ denotes CMB power spectrum.
Given CMB anisotropy data, we may estimate two point angular correlation:
\begin{eqnarray}
C(\theta)= T(\hat{\mathbf n}_1)\:T(\hat{\mathbf n}_2),
\end{eqnarray}
where $\theta=\cos^{-1}(\hat{\mathbf n}_1\cdot \hat{\mathbf n}_2)$.
Using Eq. \ref{T_expansion} and \ref{C_lm}, we may easily show that the expectation value of the correlation is given by \citep{structure_formation}:
\begin{eqnarray}
\langle C(\theta) \rangle=\sum_l \frac{2l+1}{4\pi}\,W_l\,C_l\,P_l(\cos\theta), \label{cor}
\end{eqnarray}
where $\theta$ is a separation angle, $W_l$ is the window function of the observation and $P_l$ is a Legendre polynomial. 
From Eq. \ref{cor}, we may easily see the angular correlation $C(\theta)$ and power spectrum $C_l$ possess some equivalence.

\section{Lack of angular correlation in the WMAP data}
\label{wmap}
\begin{figure}[htb!]
\centering\includegraphics[scale=.55]{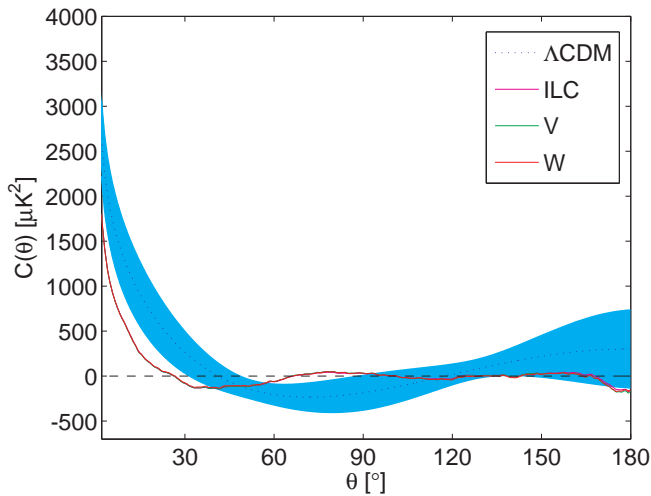}
\caption{Angular correlation of CMB anisotropy: Solid lines denote the angular correlation of WMAP data. Dotted line and shaded region denote the theoretical prediction and 1$\sigma$ ranges, as determined by Monte-Carlo simulations ($\Lambda$CDM).}
\label{C_data}
\end{figure}

In Fig. \ref{C_data}, we show the angular correlation of the WMAP 7 year data, which are estimated respectively from the WMAP team's Internal Linear Combination (ILC) map, and foreground reduced maps of V and W band. 
In the angular correlation estimation, we have excluded the foreground-contaminated region by applying the WMAP KQ75 mask, as recommended for non-Gaussianity study \citep{WMAP7:fg}.
In the same plot, we show the angular correlation of the WMAP concordance model \citep{WMAP7:Cosmology}, where the dotted line and shaded region denote the 
mean value and 1$\sigma$ ranges of Monte-Carlo simulations at V band.
%theoretical prediction and 1$\sigma$ ranges, as determined by Monte-Carlo simulations of the V band.
For simulation, we have made $10^4$ realizations with the same configuration with the WMAP data (e.g. a foreground mask, beam smoothing and instrument noise).
In order to include WMAP noise in simulation, we have subtracted one Differencing Assembly (D/A) data from another, and added it to simulations.

As shown in Fig. \ref{C_data}, there exists non-negligible discrepancy between the data and the theoretical prediction.
Most noticeably, angular correlation of WMAP data nearly vanishes at angles larger than $\sim 60^\circ$, which are previously investigated by 
\citep{correlation_COBE,WMAP1:Cosmology,correlation_Copi1,correlation_Copi2,lowl_anomalies}.
In the previous investigations, the lack of large-angle correlation has been assessed by the following statistic \citep{WMAP1:Cosmology,correlation_Copi1,correlation_Copi2,lowl_anomalies}:
\begin{eqnarray}
S_{1/2}=\int^{1/2}_{-1} \left( C(\theta) \right)^2 d(\cos\theta). \label{S12}
\end{eqnarray}
The investigation shows the $S_{1/2}$ estimated from WMAP data is anomalously low, which requires the chance $\lesssim 10^{-3}$ \citep{WMAP1:Cosmology,correlation_Copi1,correlation_Copi2,lowl_anomalies,lowl_bias}.
Besides the lack of correlation at large angles, we may see from Fig. \ref{C_data} that correlation at small angles tends to be smaller than the theoretical prediction.
Noting this, we have investigated the small-angle correlation with the following statistics:
\begin{eqnarray}
S_{\sqrt{3}/2}&=&\int^{1}_{\sqrt{3}/2} \left( C(\theta) \right)^2 d(\cos\theta),
\end{eqnarray}
where the square of the correlation is integrated over small angles ($0\le \theta\le30^\circ$).

\begin{table}[htb!]
\centering
\caption{$S$ statistics of WMAP 7 year data}
\begin{tabular}{ccccc}
\hline\hline 
  & band &  angles&value [$\mu\mathrm{K}^4$]& $p$-value\\
\hline
$S_{1/2}$ & V  &  $60^\circ\le \theta\le180^\circ$&$1.42\times 10^{3}$  & $8\times 10^{-4}$\\
$S_{1/2}$ & W  &  $60^\circ\le \theta\le180^\circ$&$1.32\times 10^{3}$  & $6\times 10^{-4}$\\
\hline
$S_{\sqrt{3}/2}$ & V  & $0^\circ\le \theta\le30^\circ$ &$2.02\times 10^{4}$  & $3.2\times 10^{-3}$\\
$S_{\sqrt{3}/2}$ & W  & $0^\circ\le \theta\le30^\circ$ &$2.03\times 10^{4}$  & $3.2\times 10^{-3}$\\
\hline
\end{tabular}
\label{pvalue}
\end{table}
In Table \ref{pvalue}, we show $S_{1/2}$ and $S_{\sqrt{3}/2}$ of the WMAP 7 year data.
Note that the slight difference between V and W band is due to the distinct beam size, and simulations are made accordingly for each band.
In the same table, we show the $p$-value, where the $p$-value denotes fractions of simulations as low as those of WMAP data.
As shown in Table \ref{pvalue}, WMAP data have unusually low values of $S_{1/2}$ and $S_{\sqrt{3}/2}$, as indicated by their $p$-value.
Note that the $p$-value of $S_{\sqrt{3}/2}$ corresponds to very high statistical significance, even though it may not be as low as that of $S_{1/2}$, 
Since $S_{\sqrt{3}/2}$ and $S_{1/2}$ corresponds to the integrated power at small and large angles respectively, we find anomalous lack of correlation at small angles in addition to large angles.

\begin{figure}[htb!]
\centering\includegraphics[scale=.48]{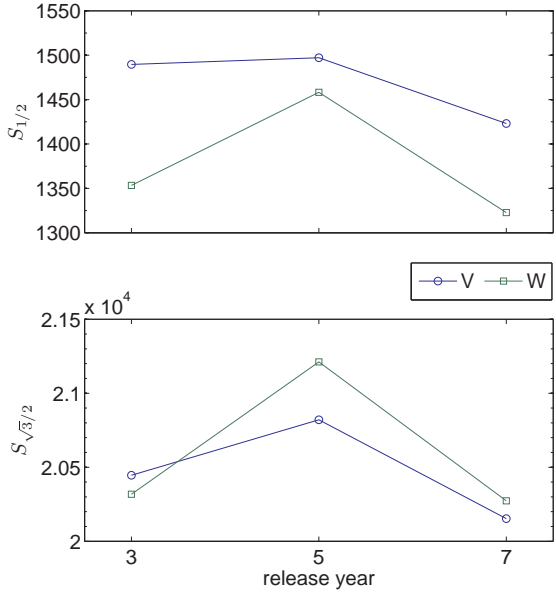}
\caption{$S$ statistics of WMAP 3, 5 and 7 year data}
\label{S_data}
\end{figure}
In Fig. \ref{S_data}, we show $S_{1/2}$ and $S_{\sqrt{3}/2}$, which are estimated from the WMAP 3, 5 and 7 year data respectively.
As shown in Fig. \ref{S_data}, the $S$ statistics of WMAP 7 year data are lowest, while WMAP 7 year data are believed to have more accurate calibration and less foreground contamination than earlier releases \citep{WMAP5:basic_result,WMAP7:basic_result,WMAP7:fg}. 
Therefore, we may not readily attribute the anomaly to calibration error or foregrounds.

We have also slightly varied the partition of $S$ within $\pm 5^\circ$. 
The pvalue of $S_{\sqrt{3}/2}$ stays the same, when the bound of the partition is set to $25^\circ \sim 32^\circ$, and
increases slightly, when the bound is $35^\circ$. 
For $S_{1/2}$, the pvalue almost stays the same, and decreases even further, when the bound of the partition is set to $62^\circ \sim 64^\circ$.
Therefore, we find our results are robust to the slight variations in the partition, and the enhancement on the statistical significance by the posteriori choice of the partition is not significant.

\section{Non-cosmological contamination}
\label{systematics}
The WMAP data contain contamination from residual galactic and extragalactic foregrounds, even though 
we have applied the conservative KQ75 mask\citep{WMAP7:fg}.
In order to investigate residual foregrounds, we have subtracted the foreground-reduced W band map from that of V band.
This difference map mainly contains residual foregrounds at V and W band maps with slight amount of CMB. Note that CMB signal is not completely cancelled out, because the beam size at V and W band differs from each other.
From the difference map $V(\mathbf n)-W(\mathbf n)$, we have obtained $S_{1/2}=0.31$ and $S_{\sqrt{3}/2}=31.36$. By comparing these values with Table \ref{pvalue}, we may see residual foregrounds at V and W band are too small to affect the correlation power of WMAP data.

There are instrument noise in the WMAP data.
Especially, 1/f noise, when coupled with WMAP scanning pattern, may result in less accurate measurement at certain angular scales  \citep{WMAP3:temperature,Detection_Light,WMAP1:processing}. 
In order to investigate the association of noise with the anomaly, we have produced noise maps of WMAP7 data by subtracting one Differencing Assembly (D/A) map from another D/A data of the same frequency channel. 
\begin{table}[htb]
\centering
\caption{the $S$ statistics of WMAP instrument noise in the unit of [$\mu\mathrm{K}^4$]}
\begin{tabular}{ccc}
\hline\hline 
data &$S_{1/2}$ & $S_{\sqrt{3}/2}$ \\
\hline
V1-V2 & 0.25 &  83.94 \\
W1-W2 &  2.49 & 587.45 \\
W1-W3 &  2.18 & 664.26 \\
W1-W4 & 2.24 & 625.27 \\
W2-W3 & 2.72 & 808.32 \\
W2-W4 &  4.39 & 764.96\\
W3-W4 & 4.39 & 764.96\\
\hline
\end{tabular}
\label{S_noise}
\end{table}
In Table \ref{S_noise}, we show $S_{1/2}$ and $S_{\sqrt{3}/2}$ estimated from the noise maps.
Comparing Table \ref{pvalue} with Table \ref{S_noise}, we may see the noise is not significant enough to cause the correlation anomalies of the the WMAP data.

\begin{figure}[htb!]
\centering\includegraphics[scale=.5]{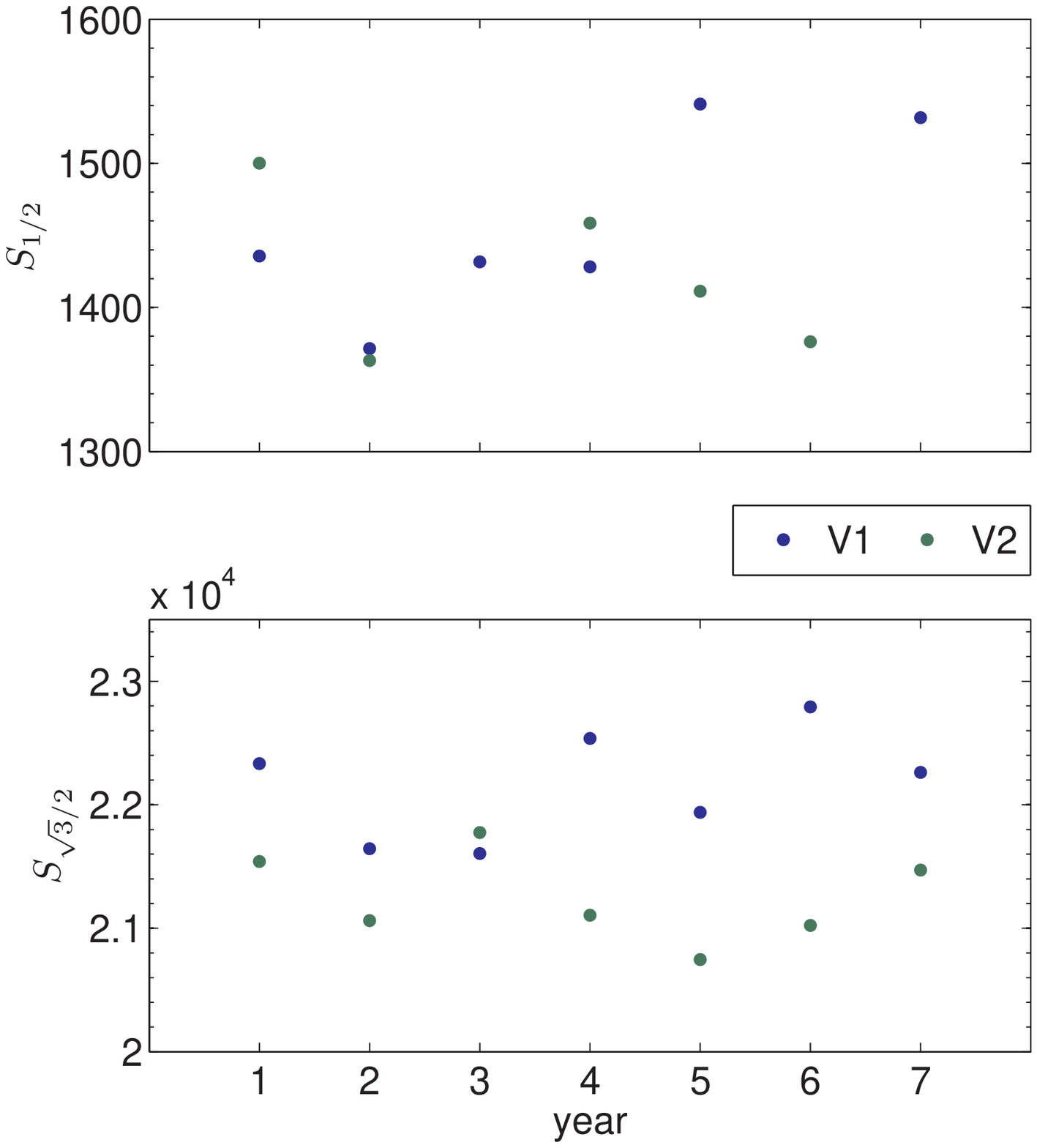}
\centering\includegraphics[scale=.5]{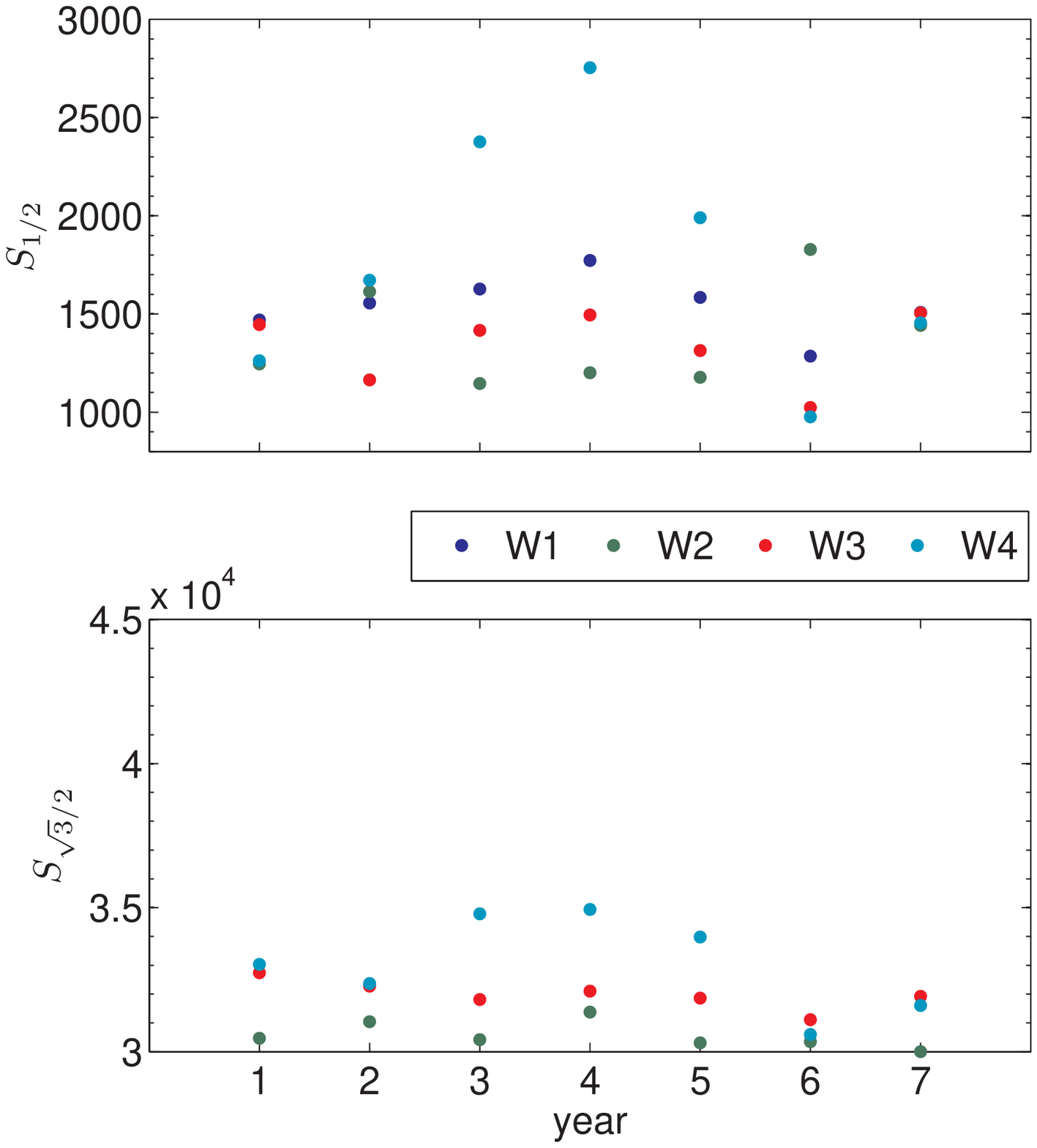}
\caption{the $S$ statistics of WMAP data at each D/A and year}
\label{S_year}
\end{figure}
In Fig. \ref{S_year}, we show the values of $S_{1/2}$ and $S_{\sqrt{3}/2}$ for each year and D/A data. As shown in Fig. \ref{S_year}, we find the anomaly is not associated with a particular D/A channel nor a year data, but present at all year and D/A channels.

\begin{figure}[htb!]
\centering\includegraphics[scale=.5]{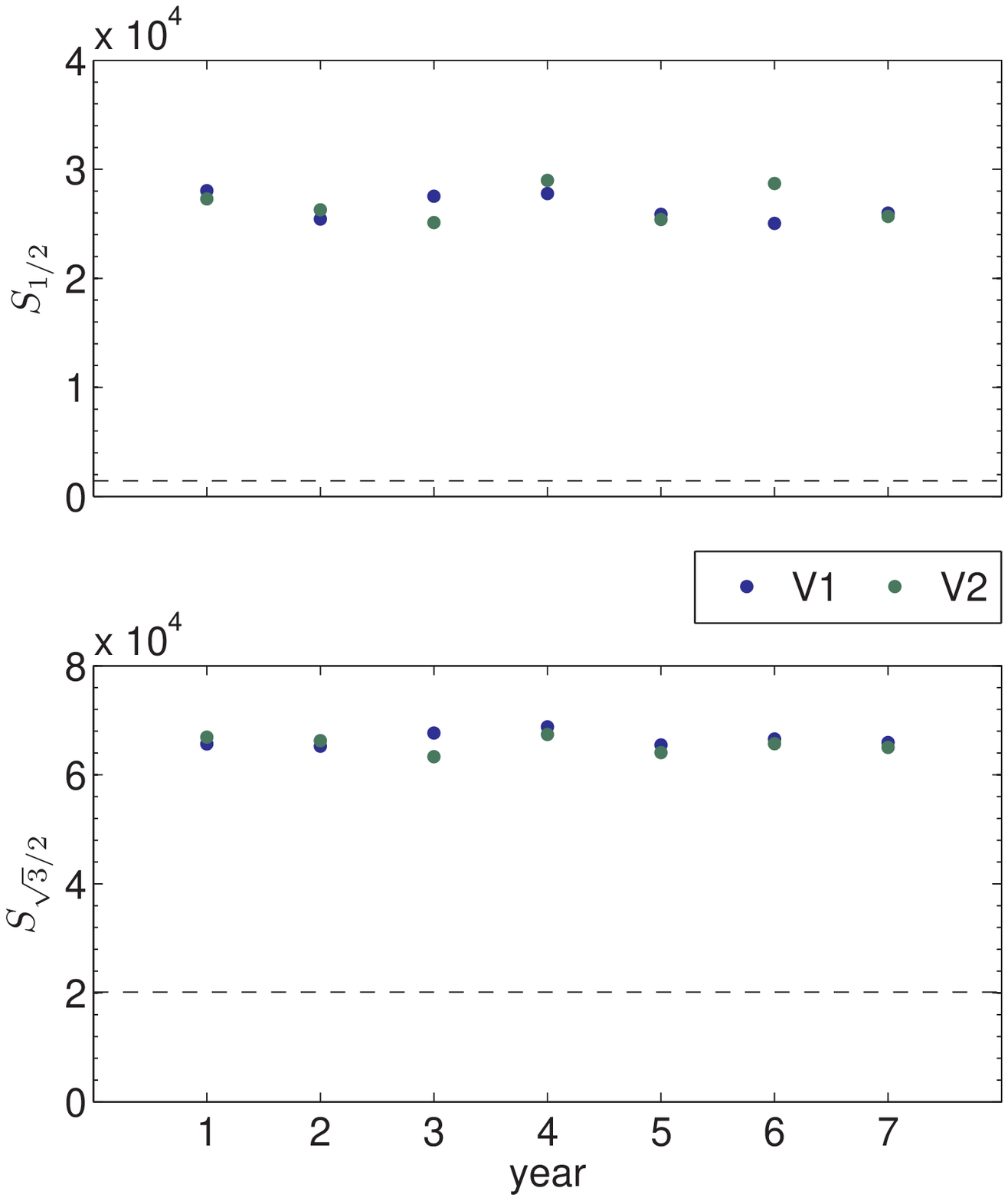}
\centering\includegraphics[scale=.5]{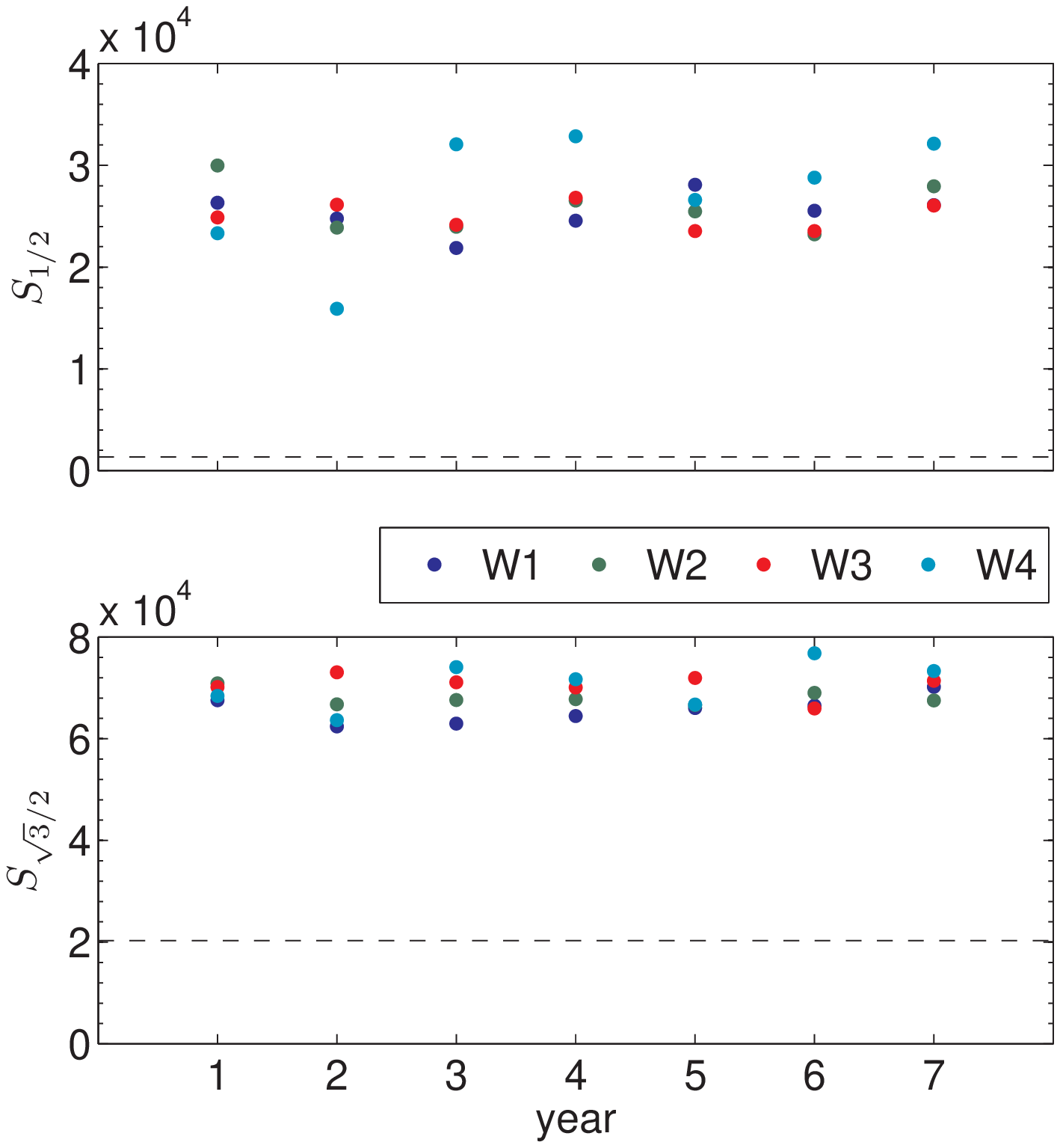}
\caption{the $S$ statistics of the simulated data produced by the WMAP team. Dashed lines show the values of WMAP data}
\label{S_sim}
\end{figure}
Besides contamination discussed above, there are other sources of contamination such as sidelobe pickup and so on.
In order to investigate these effects, we have investigated simulations produced by the WMAP team.
According to the WMAP team, time-ordered data (TOD) have been simulated with realistic noise, thermal drifts in instrument gains and baselines,
smearing of the sky signal due to finite integration time, transmission imbalance and far-sidelobe beam pickup.
Using the same data pipeline used for real data, the WMAP team have processed simulated TOD, and produced maps for each D/A and each year. 
From the simulated maps, we have estimated $S_{1/2}$ and $S_{\sqrt{3}/2}$, which are plotted in Fig. \ref{S_sim}.
As shown in Fig. \ref{S_sim}, $S$ statistics of simulated data are significantly higher than those of WMAP data.
Therefore, the anomaly may be produced by unknown systematics or indeed cosmological.

\section{Odd multipole preference in CMB power spectrum data}
\label{odd}
Without loss of generality, we may consider CMB anisotropy field as the sum of even and odd parity functions:
\begin{eqnarray} 
T(\hat{\mathbf n})=T^+(\hat{\mathbf n})+T^-(\hat{\mathbf n}),  
\end{eqnarray}
where
\begin{eqnarray} 
T^+(\hat{\mathbf n})&=&\frac{T(\hat{\mathbf n})+T(-\hat{\mathbf n})}{2},\\
T^-(\hat{\mathbf n})&=&\frac{T(\hat{\mathbf n})-T(-\hat{\mathbf n})}{2}.
\end{eqnarray}
Using Eq. \ref{T_expansion} and the parity property of spherical harmonics $Y_{lm}(\hat{\mathbf n})=(-1)^l\,Y_{lm}(-\hat{\mathbf n})$ \citep{Arfken},
we may show
\begin{eqnarray} 
T^+(\hat{\mathbf n})&=&\sum_{lm} a_{lm}\,Y_{lm}(\hat{\mathbf n})\,\cos^2\left(\frac{l\pi}{2}\right), \label{T_even}\\
T^-(\hat{\mathbf n})&=&\sum_{lm} a_{lm}\,Y_{lm}(\hat{\mathbf n})\,\sin^2\left(\frac{l\pi}{2}\right). \label{T_odd}
\end{eqnarray}
Obviously, the power spectrum of even and odd multipoles are associated with $T^{+}(\hat{\mathbf n})$ and $T^{-}(\hat{\mathbf n})$ respectively.
Given the $\Lambda$CDM model, we do not expect any features distinct between even and odd multipoles. However, 
there are reports on anomalous power excess (deficit) at odd (even) multipoles data ($2\le l\le 22$), which have been dubbed as `odd-parity preference'   \citep{odd,odd_origin,odd_bolpol}. 

Angular power spectrum and angular correlation possess some equivalence.
Noting this, we have investigated the association of the odd-parity preference with the lack of large-angle correlation.
Using Eq. \ref{cor} with the Sach plateau approximation (i.e $l(l+1)\,C_l/2\pi\sim \mathrm{const}$), we find the expectation value of angular correlation is given by:
\begin{eqnarray}
C(\theta)&=&\sum_l \frac{2l+1}{4\pi}\,W_l\,C_l\,P_l(\cos\theta)\label{C_lowl}\\
&=&\sum_l \frac{l(l+1)\,C_l}{2\pi} \,\frac{2l+1}{2l(l+1)}\,W_l\,\,P_l(\cos\theta)\nonumber\\
&\approx&  \alpha \sum^{l_0}_{l} \frac{2l+1}{2l(l+1)}\,W_l\,\,P_l(\cos\theta)\nonumber\\
&&+\sum_{l=l_0+1} C_l\,\frac{2l+1}{4\pi}\,W_l\,P_l(\cos\theta),\nonumber
\end{eqnarray}
% \begin{eqnarray}
% C(\theta)&=&\sum_l \frac{2l+1}{4\pi}\,W_l\,C_l\,P_l(\cos\theta)\label{C_lowl}\\
% &=&\sum_l \frac{l(l+1)\,C_l}{2\pi} \,\frac{2l+1}{2l(l+1)}\,W_l\,\,P_l(\cos\theta)\nonumber\\
% &\approx&  \alpha \sum^{l_0}_{l} \frac{2l+1}{2l(l+1)}\,W_l\,\,P_l(\cos\theta)+\sum_{l=l_0+1} C_l\,\frac{2l+1}{4\pi}\,W_l\,P_l(\cos\theta),\nonumber
% \end{eqnarray}
where $\alpha$ is some positive constant and $l_0$ is a low multipole number, within which the Sach plateau approximation is valid.
As discussed above, there exists the odd multipole preference at low multipole ($2\le l\le22$).
Considering the odd multipole preference, we may show the angular correlation is given by:
% \begin{eqnarray}
% C(\theta)\approx\alpha(1-\varepsilon)\,F(\theta)+\alpha(1+\varepsilon)\,G(\theta)+\sum_{l=23} C_l\,\frac{2l+1}{4\pi}\,W_l\,P_l(\cos\theta),\label{C_FG}
% \end{eqnarray}
\begin{eqnarray}
\lefteqn{C(\theta)}\label{C_FG}\\
&\approx& \alpha(1-\varepsilon)\,F(\theta)+\alpha(1+\varepsilon)\,G(\theta)+\sum_{l=23} C_l\,\frac{2l+1}{4\pi}\,W_l\,P_l(\cos\theta),\nonumber
\end{eqnarray}
where $\varepsilon$ is some positive constants, and
\begin{eqnarray*}
F(\theta)&=&\sum^{22}_{l}\,\frac{2l+1}{2l(l+1)}\,W_l\,\,P_l(\cos\theta)\:\cos^2\left(\frac{l\pi}{2}\right),\\
G(\theta)&=&\sum^{22}_{l}\,\frac{2l+1}{2l(l+1)}\,W_l\,\,P_l(\cos\theta)\:\sin^2\left(\frac{l\pi}{2}\right).
\end{eqnarray*}
In Eq. \ref{C_FG}, $\alpha\varepsilon(-F(\theta)+G(\theta))$ corresponds to the deviation from the standard model, due to the odd multipole preference ($2\le l\le22$).

\begin{figure}[htb!]
\centering\includegraphics[scale=.54]{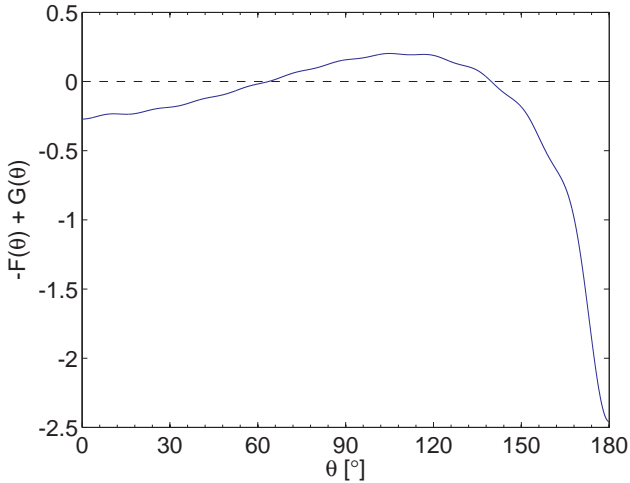}
\caption{the effect of the odd multipole preference on the correlation}
\label{W}
\end{figure}
\begin{figure}[htb!]
\centering\includegraphics[scale=.54]{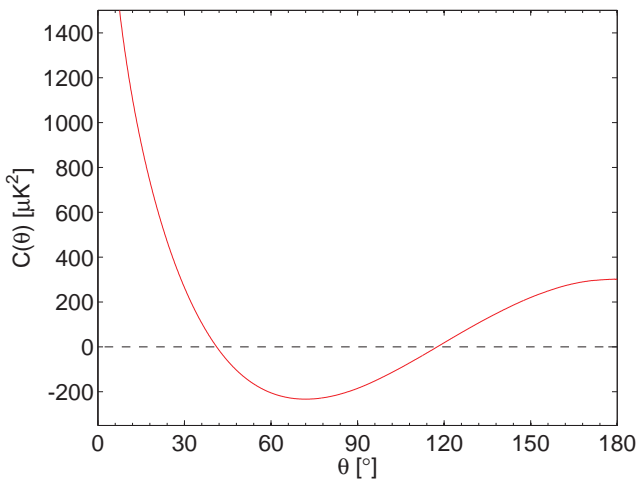}
\caption{the angular correlation without odd-parity preference (i.e. Eq. \ref{cor})}
\label{C_theory}
\end{figure}
In Fig. \ref{W} and \ref{C_theory}, we show $-F(\theta)+G(\theta)$ and the angular correlation of the standard model (i.e. $\varepsilon=0$). 
Let us consider the intervals $60^\circ\le \theta\le 120^\circ$ and $120^\circ\le \theta\le 180^\circ$, which are associated with the statistic $S_{1/2}$.
At the interval $60^\circ\le \theta\le 120^\circ$, the angular correlation has negative values, while the deviation $\alpha\,\varepsilon(-F(\theta)+G(\theta))$ is positive. 
At the interval $120^\circ\le \theta\le 180^\circ$, the angular correlation has positive values, while the deviation $\alpha\,\varepsilon(-F(\theta)+G(\theta))$ is negative. 
Therefore, we find
\begin{eqnarray}
(\left.C(\theta)\right|_{\varepsilon>0})^2<(\left.C(\theta)\right|_{\varepsilon=0})^2\qquad(60^\circ\le \theta\le 180^\circ). \label{C_ne}
\end{eqnarray}
From Eq. \ref{C_ne}, we may see the odd-parity preference (i.e. $\epsilon>0$) leads to lack of large-angle correlation power.

\begin{figure}[htb!]
\centering\includegraphics[scale=.54]{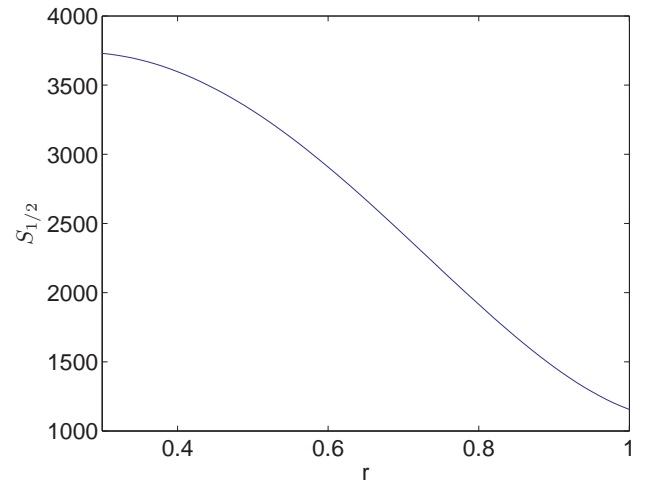}
\caption{$S_{1/2}$ of the WMAP team's Internal Linear Combination (ILC) map, where the octupole components is multiplied by the suppression factor $r$.}
\label{C_l3}
\end{figure}

We like to emphasize that the lack of large-correlation is associated with the odd-parity preference at low multipoles (i.e. power excess at even multipoles and power deficit at odd multipoles). On the other hand, simple suppression of overall low multipole power does not necessarily leads to the lack of large-angle correlation.
For instance, suppressing octupole power, which mitigates the odd-parity preference, rather increases the large-angle correlation power.
In Fig. \ref{C_l3}, we show $S_{1/2}$ of the WMAP team's Internal Linear Combination (ILC) map, where we have multiplied the suppression factor $r$ to the quadrupole component of the map. From Fig. \ref{C_l3}, we may see that the large-angle correlation power increases, as the octupole component are more suppressed.

\section{Possible Cosmological origin}
\label{cosmic}
As discussed previously, we have not found a definite non-cosmological cause of the discussed anomaly.
Therefore, in this section, we are going to consider possible cosmological origins.
Since primordial fluctuations, which were once on sub-Planckian scales, are stretched to the observable scales by inflation, trans-Planckian effects may leave imprints on a primordial power spectrum \citep{Inflation_Planckian_problem,Inflation_Planckian_spectra,Inflation_Planckian_note,Inflation_Planckian_estimate,CMB_Planckian_signature,WMAP_oscillation,Inflation_Planckian,Inflation_initial}.
Though trans-Planckian imprints are highly model-dependent \citep{Planckian_Astrophysics,CMB_Planckian_observation}, most of the models predict oscillatory features in primordial power spectrum \citep{Inflation,Inflation_Planckian_problem,Inflation_Planckian_spectra,Inflation_Planckian_note,Inflation_Planckian_estimate,CMB_Planckian_signature,WMAP_oscillation,Inflation_Planckian,Inflation_initial,Planckian_Astrophysics,CMB_Planckian_observation,WMAP3:parameter}. 
Given certain oscillatory features in primordial power spectrum, the trans-Planckian effects may produce the observed odd parity preference of CMB power spectrum.

However, reconstruction of primordial power spectrum and investigation on features have not found a strong evidence for features in the primordial powerspectrum \citep{WMAP7:powerspectra,WMAP5:Cosmology,WMAP7:Cosmology,WMAP3:parameter,power_recon,power_svd,power_features}.
Therefore, we are going to consider what condition the observed odd-parity preference impose on primordial fluctuation, if a primordial power spectrum is a featureless power-law spectrum. Decomposition coefficients are related to primordial perturbation as follows:
\begin{eqnarray}
a_{lm}&=&4\pi (-\imath)^l \int \frac{d^3\mathbf k}{(2\pi)^3} \Phi(\mathbf k)\,g_{l}(k)\,Y^*_{lm}(\hat {\mathbf k}),\label{alm}
\end{eqnarray} 
where $\Phi(\mathbf k)$ is primordial perturbation in Fourier space, and $g_{l}(k)$ is a radiation transfer function.
Using Eq. \ref{alm}, we may show the decomposition coefficients of CMB anisotropy are given by: 
\begin{eqnarray*}
a_{lm}&=&\frac{(-\imath)^l}{2\pi^2} \int\limits^{\infty}_0 dk  \int\limits^{\pi}_0 d\theta_{\mathbf k} \sin\theta_{\mathbf k}\\
&\times&\int\limits^{\pi}_0 d\phi_{\mathbf k}\,g_{l}(k) Y^*_{lm}(\hat {\mathbf k})\left(\Phi(\mathbf k)+(-1)^l \Phi^*(\mathbf k)\right),
\end{eqnarray*}
where we used the reality condition $\Phi(-\mathbf k)= \Phi^*(\mathbf k)$ and $Y_{lm}(\hat{-\mathbf n})=(-1)^l\,Y_{lm}(\hat{\mathbf n})$.
Using Eq. \ref{alm2},  it is trivial to show, for the odd number multipoles $l=2n-1$,
\begin{eqnarray}
\lefteqn{a_{lm}=}\label{alm2}\\
&-&\frac{(-\imath)^{l-1}}{\pi^2} \int\limits^{\infty}_0 dk  \int\limits^{\pi}_0 d\theta_{\mathbf k} \sin\theta_{\mathbf k}\int\limits^{\pi}_0 d\phi_{\mathbf k}\,g_{l}(k) Y^*_{lm}(\hat {\mathbf k})\,\mathrm{Im}[\Phi(\mathbf k)],\nonumber
\end{eqnarray}
and, for even number multipoles $l=2n$,
\begin{eqnarray}
\lefteqn{a_{lm}=}\label{alm3}\\
&&\frac{(-\imath)^l}{\pi^2} \int\limits^{\infty}_0 dk  \int\limits^{\pi}_0 d\theta_{\mathbf k} \sin\theta_{\mathbf k}\int\limits^{\pi}_0 d\phi_{\mathbf k}\,g_{l}(k) Y^*_{lm}(\hat {\mathbf k})\,\mathrm{Re}[\Phi(\mathbf k)].\nonumber
\end{eqnarray}
It should be noted that the above equations are simple reformulation of Eq. \ref{alm}, and exactly equal to them.
From Eq. \ref{alm2} and \ref{alm3}, we may see that the observed odd-parity preference might be produced, provided
\begin{eqnarray}
|\mathrm{Re} [\Phi(\mathbf k)]|\ll|\mathrm{Im} [\Phi(\mathbf k)]|\;\;\;(k\lesssim 22/\eta_0),\label{primordial_odd} 
\end{eqnarray}
where $\eta_0$ is the present conformal time.
Taking into account the reality condition $\Phi(-\mathbf k)= \Phi^*(\mathbf k)$, we may show primordial perturbation in real space is given by:
\begin{eqnarray}
\Phi(\mathbf x)&=&2\int\limits^{\infty}_0 dk  \int\limits^{\pi}_0 d\theta_{\mathbf k} \sin\theta_{\mathbf k}\label{Phi_real}\\
&\times&\int\limits^{\pi}_0 d\phi_{\mathbf k}\left(\mathrm{Re}[\Phi(\mathbf k)]\cos(\mathbf k\cdot \mathbf x)-\mathrm{Im}[\Phi(\mathbf k)]\sin(\mathbf k\cdot \mathbf x)\right).\nonumber 
\end{eqnarray}
Noting Eq. \ref{primordial_odd} and \ref{Phi_real}, we find our primordial Universe may possess odd-parity preference on large scales ($2/\eta_0 \lesssim k\lesssim 22/\eta_0$).
This explanation requires the violation of the large-scale translational invariance, putting us at a special place in the Universe.
However, it is not in direct conflict with the current data on observable Universe (i.e. WMAP CMB data), and the invalidity of the Copernican Principle such as our living near the center of void had been already proposed in different context \citep{Void_DE,Void_SN}.

Independently, there exist some theoretical models, which predict a parity-odd local Universe \citep{Parity_odd_Universe,CP_violation}.
In the models, some level of non-zero TB and EB correlations are predicted as well \citep{Parity_odd_Universe}.

Depending on the type of cosmological origins, distinct anomalies are predicted in polarization power spectrum and correlations (e.g. TB, EB). 
Therefore, polarization maps of large-sky coverage (i.e. low multipoles) will allow us to remove degeneracy and
figure out a cosmological origin, provided the odd parity preference is indeed cosmological.

\section{Discussion}
\label{discussion}
We have investigated the angular correlation in the recent CMB data.
In addition to the well-known correlation anomaly at large angles, we find the lack of correlation at small angles with high statistical significance.

In the angular power spectrum of WMAP data, there exist anomalous odd-parity preference at low multipoles \citep{Universe_odd,odd,odd_origin,odd_bolpol,odd_tension}.
Angular power spectrum and angular correlation possess some equivalence.
Noting this, we have investigated the association between the lack of correlation and the odd-parity preference.
We find that the odd-parity preference is, in fact, a phenomenological origin of the correlation anomaly \citep{odd,odd_origin,odd_bolpol}.

We have investigated non-cosmological contamination and the WMAP team's simulated data. However, we have not found a definite cause. 
The Planck surveyor data possesses wide frequency coverage and systematics distinct from the WMAP. Therefore, it may allow us to resolve its origin. 
Most of all, Planck polarization data, which have low noise and large sky coverage, will greatly help us to understand the underlying origin of the anomaly.

\section{Acknowledgments}
We are grateful to the anonymous referee for helpful comments and suggestions.
We acknowledge the use of the Legacy Archive for Microwave Background Data Analysis (LAMBDA).
Our data analysis made the use of HEALPix \citep{HEALPix:Primer,HEALPix:framework} and SpICE \citep{spice1,spice2}.  
This work is supported in part by Danmarks Grundforskningsfond, which allowed the establishment of the Danish Discovery Center.
This work is supported by FNU grant 272-06-0417, 272-07-0528 and 21-04-0355. 

\bibliographystyle{plainnat}
\bibliography{/home/tac/jkim/Documents/bibliography}
\end{document}